# Spatial Awareness of a Bacterial Swarm

Harshitha S. Kotian[1][2], Shalini Harkar[2], Shubham Joge[3], Ayushi Mishra[3], Amith Zafal[1], Varsha Singh[3], Manoj M. Varma[1][2]

[1]Centre for Nano Science and Engineering, Indian Institute of Science, Bangalore, India
[2] Robert Bosch Centre for Cyber Physical Systems, Indian Institute of Science, Bangalore, India
[3]Molecular Reproduction, Development and Genetics, Indian Institute of Science, Bangalore, India

**Abstract**

Bacteria are perhaps the simplest living systems capable of complex behaviour involving sensing and coherent, collective behaviour an example of which is the phenomena of swarming on agar surfaces. Two fundamental questions in bacterial swarming is how the information gathered by individual members of the swarm is shared across the swarm leading to coordinated swarm behaviour and what specific advantages does membership of the swarm provide its members in learning about their environment. In this article, we show a remarkable example of the collective advantage of a bacterial swarm which enables it to sense inert obstacles along its path. Agent based computational model of swarming revealed that independent individual behaviour in response to a two-component signalling mechanism could produce such behaviour. This is striking because independent individual behaviour without any explicit communication between agents was found to be sufficient for the swarm to effectively compute the gradient of signalling molecule concentration across the swarm and respond to it.

**Introduction**

Bacteria are perhaps the simplest living systems capable of complex behaviour involving sensing and coherent, collective behaviour an example of which is the phenomena of swarming, where bacteria colonize a solid surface (typically, nutrient loaded agar) in geometric patterns characteristic of different species (Ben-Jacob, 1997; Kearns, 2010). A natural and fascinating question in this regard is how the information gathered by individual members of the swarm through sensing their respective local environments is shared across the swarm leading to coordinated swarm behaviour. Specifically, does membership of the swarm provide its members any advantage in learning more about their environment? For instance, are they able to extract more information about their surroundings collectively than acting alone? Swarming involves several, possibly collective, decision making steps such as quorum sensing (Daniels et al., 2004). The swarming patterns produced by the Gram-negative bacteria *Pseudomonas aeruginosa* (PA) are special due to the presence of long straight segments (tendrils) in its swarming pattern [Figure 1 a)]. The swarming pattern produced by PA is unique in two major aspects. Firstly, the typical tendril length (~ cm) is nearly 4 orders of magnitude compared to the micron-sized individual swarm member. Considering the large stochastic fluctuations likely to be experienced by a micron sized individual member, it is remarkable that collective motion of these individuals produces such long straight segments. The long directional persistence leading to this highly anisotropic collective motion is worthy of detailed mathematical modelling. Indeed, there have been several attempts to model the swarming pattern of PA based on a highly mechanistic multiscale model(Du et al., 2011), as a population dispersal phenomenon using spatial kernels(Deng et al., 2014) and based on Marangoni forces(Du et al., 2012). However, so far none of these have succeeded in producing patterns with similar branching statistics similar to the one seen in Figure 1 a) [SI Section1 Figure S2]. While generic models such as the Vicsek model [(Vicsek et al., 1995)] can explain large directional persistence collections of stochastically moving particles, one would like to have a more detailed model involving experimentally accessible parameters of the system which produces statistically identical patterns as seen in the experimental system. While this is an open problem, in this paper we focus on a different shortcoming of the existing models which is that these models have primarily been employed to study the pattern formation and do not describe other aspects of swarming such as the question we posed earlier concerning sensing of the environment by swarms.

The tendrils produced by swarming PA are typically of the order of a centimetre with a width of 4-5 mm and their motion during swarming can be easily tracked with time-lapse imaging using regular digital cameras. This enables us to study the motion of the tendrils and their response to various environmental perturbations with relative ease compared to isotropically swarming bacteria (Kearns & Losick, 2003) and other bacterial species which form dense fractal-like patterns (Ben-Jacob, 1997). The motion of single bacterial cells on the agar surface can also be observed using GFP expressing bacteria leading to the generation of motility data spanning multiple spatial (microns to cm) and temporal (seconds to days) scales. Such multi-scale imaging data helps to link individual behaviour to the collective behaviour of the swarm. The observation and analysis of the response of the tendrils to perturbations introduced into the swarming medium (agar) reveals a remarkable ability of PA swarm to sense its spatial environment and respond to the presence of co-swarming sister tendrils (other tendrils from the same swarm initiating colony) [Figure 1 b)], approaching boundary of the petri-plate and even inert obstacles (Poly DiMethyl Siloxane (PDMS) and glass objects), several millimetres away along the swarming path. The ability to sense inert objects at millimetre scale distances is a particularly striking example of the spatial awareness of the PA swarm considering that the swarm is able to sense objects at distances three orders of magnitude (mm scale) further than body length of the individual members (micron scale). This fact is indicative of the collective advantage provided by the swarm as it is inconceivable that individual bacterium can sense inert objects at such long distances. We emphasize that the swarm is spatially aware of the object, i.e. the object does not actively secrete any molecules which are sensed by the bacterium.

The observation of spatial awareness by the swarm leads to the question of causative mechanisms. Using a continuous, fluid dynamic model involving secretion of a signalling molecule by the bacteria comprising the swarm, we show that a concentration gradient of the signalling molecule emerges within the swarm. This leads to two possible scenarios, namely, one in which the bacteria in different locations within the swarm behave differently leading to the collective response, or the other where the global gradient is implicitly computed by the swarm and to which it responds to. The latter scenario requires possibly complex information exchange within the swarm. To explore this further, we developed a multi-agent model, based on attractive-repulsive interaction arising purely from local information of the concentration of signalling molecules, which was able to replicate the spontaneous retraction (reversal of direction) of the swarm from an inert boundary. While this model is not yet a comprehensive representation of the swarming phenomenon, the ability to reproduce retraction suggests that the remarkable examples of spatial awareness seen in the PA swarm may not require active exchange of information between agents.

**Results**

**General aspects of swarming pattern**

The PA swarming system is interesting due to the highly specialized branching pattern [Figure 1 a)] and the remarkable spatial awareness and response of the swarm to its environment. A short survey of swarming patterns by *Paenibacillus dendritiformis, Paenibacillus vortex*, *Escherichia coli, Proteus mirabilis , Rhizobium etli, Serratia marcescens and Salmonella typhimurium* (Verstraeten et al.,

2008)[SI Section 1 Figure S1] species of bacteria suggests that the PA swarming pattern occupies a unique parameter space relative to the other swarm patterns. Many bacterial species such as *E. coli* and *P. mirabilis* produce dense patterns which expand isotropically from the point of inoculation. *Bacillus subtilis* (Fujikawa and Matsushita 1989) *and Paenibacillus dendritiformis* under some conditions produce fractal like patterns which can be described by DLA models (Ben-Jacob, 1997). In contrast, PA swarms expand in straight segments (tendrils) from the inoculated region. The expansion of each tendril is highly anisotropic along a straight line with constant speed [SI Section 1 Figure S3]. The overall pattern is characterized by a robust statistical distribution of branch lengths and divergence angles [SI Section 1 Figure S2].

**Awareness of the Presence of Sister Tendrils**

The swarming tendrils display self-avoidance (Caiazza et al. 2005)(Tremblay et al., 2007). This behaviour is also seen in growing *Bacillus subtilis* tendrils (James et al., 2009) and in *P.. dendritiformis* by Avraham Be'er et al (Be'er et al., 2009) who used the term sibling rivalry to describe this behaviour. In this case swarm fronts emerging from two locations in the same swarm plate lead to the formation of a zone of inhibition unpopulated by either of the advancing swarm fronts which can be viewed as a form of self-avoidance. However, there is a crucial difference between the self-avoidance seen in the PA system compared with the *Paenibacillus* system. As (Be'er et al., 2010) showed the zone of inhibition is due to mutually lethal secretions produced by the swarming *P. dendritiformis* resulting in the death of bacteria in the region between the advancing fronts producing the zone of inhibition. In the case of PA, the advancing sister tendrils sense each other and typically one of them retracts back (complete reversal) or changes its direction. [SI Section 2] This is a form of true self-avoidance and not a consequence of lethal secretions killing off the bacteria in the advancing front (James et al., 2009). In other words, the bacteria on one of the advancing tendrils sense the presence of another tendril advancing towards it and initiate changes in the motility profile which result in a change in the swarming direction. Thus, the self-avoidance phenomena in PA is significantly more complex and requires coordination far more in extent to that described previously in the case of *P. dendritiformis* and requires deeper study. We do not yet have a quantitatively accurate dynamical model of swarm direction reversal associated with the sensing of sister tendrils. However, the continuous model, based on sensing of the concentration gradient of a signalling molecule, presented later in the article is able to explain certain features such as the expected change in direction and why generally only one tendril retracts.

Another behaviour related to intra-species sensing is the avoidance of non-swarming mutant strains by advancing tendrils of swarming strains as shown in Figure 1 c) [SI Section 2 Figure S5 and video].The flgM mutant of PA is nonswarming due to the absence of flagella, yet it induces avoidance response in wild type PA. Thus this mutant likely indicates its presence, through secretions, sensed by the wildtype swarm which subsequently changes the direction of the advancing tendril.

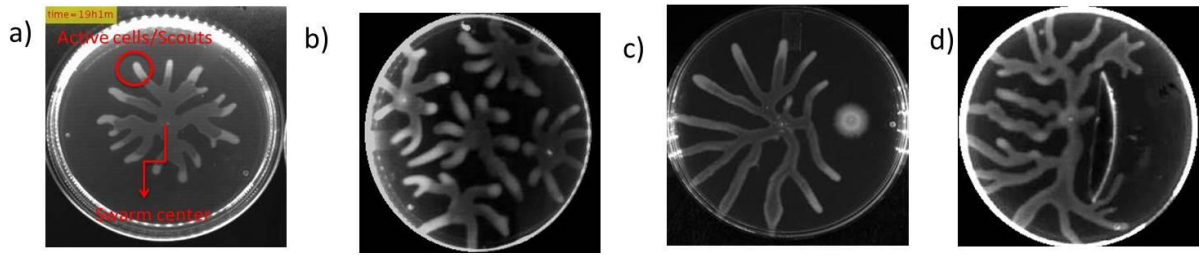

*Figure 1., a) Representative labelled picture of a PA swarm, b) Avoidance behaviour among multiple colonies of wild type on the same petri-plate, c) Avoidance exhibited by the swarm colony of the wild type in the presence of flgM mutant, d) Avoidance exhibited by the swarm colony in the presence of a passive obstacle made of PDMS.*

### Awareness of the Presence of Petri-plate Boundary

As the tendrils approach the edge of the petri-plate, we often observe a complete reversal of swarming direction. Closer observation reveals that this effect happens most likely because one of the tendrils reaches the edge of the petri-plate and releases the swarming bacteria which move along the gap between the agar and the petri-plate wall. The signalling molecule(s) from these set of bacteria are sensed by the other tendrils which have not yet reached the edge and causes direction reversal in response to this sensory information. There is further support for this hypothesis as the advancing tendrils reverse their directions only after at least one of the tendrils has reached the edge of the petri-plate [SI Section 3 for video].

### Awareness of the Presence of Inert Objects

The examples of spatial awareness presented above are all related by the fact that the target for spatial awareness was of biological origin be it the tendrils of its own colony or that of a sister colony. The sensing mechanism in this case could be hypothesized to arise out of signalling molecules secreted by the targets. We investigated if the advancing tendrils could sense the presence of inert obstacles which would not secrete signalling molecules. Interestingly, we found strong evidence of the advancing tendril detecting the presence of the obstacle and changing its direction as it approached the obstacle. The bulk of our studies were conducted with the inert polymeric material PDMS. However, in order to check the material dependence of the obstacle, we also conducted more limited studies with obstacles made of glass [SI Section 4 Figure S6]. We did rigorous statistical analysis of experimental data to quantify the detection of inert obstacles by advancing swarm tendrils. We analysed around 120 experiments involving sensing of inert objects of various shapes with negative control (no obstacle) and positive control (flgM mutant) which causes the wild type strain to avoid it as described earlier. The positive control clearly shows that the swarming pattern is perturbed by the presence of the flagellar mutant. For the inert obstacle experiments we first evaluated the possibility that the perturbation seen in the swarming pattern indeed arises due to the presence of the obstacle. This was done by computationally scanning the obstacle shape across the image and finding the best fit position. A unique fit at the exact obstacle position or the best fit being around the obstacle position supports the hypothesis that the perturbation is due to the obstacle and not due to the natural stochasticity in the swarming pattern. This analysis revealed that about 80% of experiments, a significant majority, indicate successful detection of the inert obstacle by the swarm from as far as 5mm [see SI Section 4 for video]. The detection of obstacles was also observed with change in nutrient media (From PGM to M9) as with change of obstacle material. [See SI section 4 for videos]

In addition to this analysis, we quantified the asymmetry in the swarming pattern. This is motivated by the fact that the branch distribution in the case of a negative control can be assumed to be unbiased hence symmetric. However, the presence of an inert obstacle or the positive control creates asymmetry in the swarming pattern. To factor out the asymmetry due to the presence of the obstacles itself (as opposed to the perturbation due to the obstacle) we digitally replicated the image of the obstacle (inert or positive control) in all the four quadrants and subtract the obstacle region from the original image. The pattern obtained after subtracting the digitally created obstacles is used to calculate the variance of the swarm coverage in each quadrant. Both positive control and the experiments on inert obstacles showed significantly larger inter-quadrant variance indicating strong perturbation of the baseline swarming pattern. More details of the data analysis are provided in the SI text [Section 5]. The sensing of inert obstacles from a distance represents the strongest evidence of the advantages conferred by membership of the swarm as it is inconceivable that an individual bacterium can detect an inert object from such a long distance. In the subsequent section we propose a unified model to explain all these examples of spatial awareness in PA.

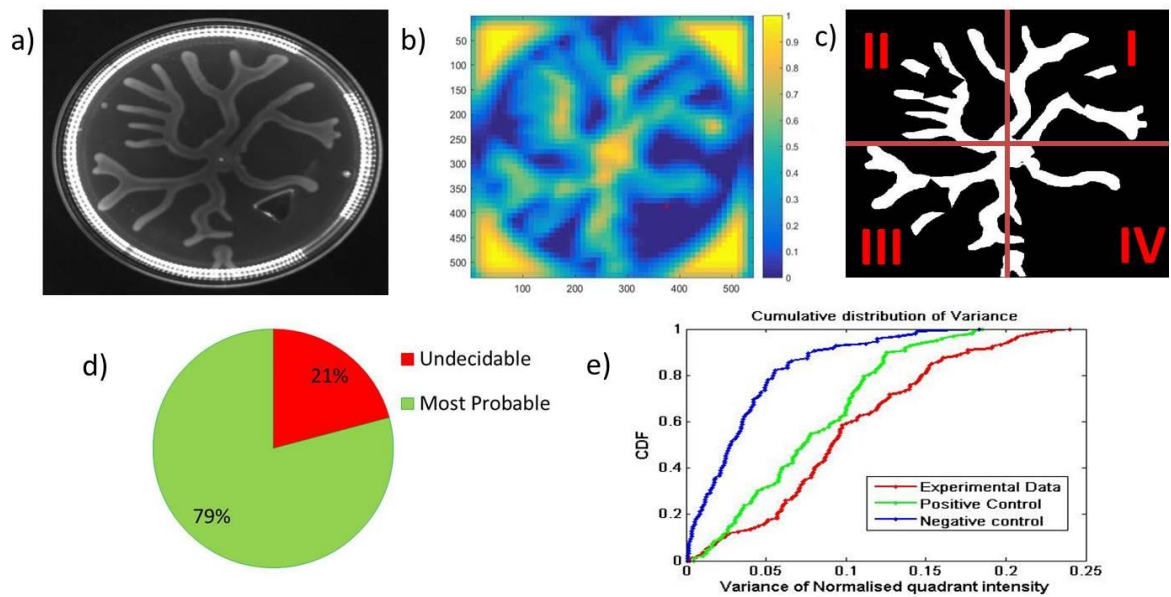

*Figure. 2, a) Swarm plate with PDMS obstacle, b) Heat map showing available positions of the given obstacle to occupy ( blue: Obstacle will not overlap with the pattern , yellow: Obstacle overlaps with the pattern) ,c) Thresholded image pattern obtained by subtracting the obstacle for asymmetry analysis (See SI Section 5 for more details), d) Statistics of different types of perturbations to the pattern due to presence of the obstacle (Most probable: Patterns encircle the obstacle while avoiding it, Undecidable: Patterns have restricted growth and/ or are unable to indicate the effect of the obstacle),  e) Cumulative Distribution Function of inter-quadrant variance in the distribution of the area occupied by the swarming colony*

**Towards a Mathematical Model of Spatial Awareness in the PA Swarm**

The main idea of the mathematical model is that the individual bacterium and consequently the swarm can sense a signalling molecule (or a cocktail of molecules) to decide its direction. For the purpose of this model it is enough to consider a single molecule even though in practice several molecules may be involved. We perform quantitative modelling to understand a) coarse-grained dynamics of swarm direction changes induced by perturbations of the various kinds mentioned above and b) what specific advantages are available to the individual members of the swarm by virtue of membership in the swarm. The requirement of a signalling molecule leading to self-avoidance of sister tendrils as well as the other instances of spatial awareness mentioned in this article is well supported by previous reports describing the role of rhamnolipids (RL) in PA swarming [(Caiazza et al., 2005)(Tremblay et al., 2007)]. We suppose that the advancing swarm tip contains active (motile) bacteria which secrete signalling molecules at a specified rate f. The signalling molecules diffuse with diffusion constant D. The governing equations then become

$$\frac{\partial C(r,t)}{\partial t} = D\nabla^2 C(r,t) + f \quad (1)$$

With f = 0 in the region outside the swarm tip. Inert obstacles or petri-plate edges are represented by reflecting (non-diffusive) boundaries. Parameters such as the production rate and diffusion constant of the signalling molecule, number of bacteria in a swarm are required to draw meaningful inferences from this model. Although these parameters have not been explicitly measured, we can estimate these from experiments if available, or by order of magnitude calculations [See SI Section 6 for more details]

Simulations proceed by initializing a circular disk of fixed radius representing the advancing swarm tip with bacteria secreting the signalling molecule. The concentration field of the signalling molecule is calculated using Eq. (1) above over the entire region and the "swarm tip" is advanced to a new position representing the motion of the swarming tendril. The concentration field is updated and the process continues. Firstly, we see that for the best estimates we have for the model parameters, a gradient in the order of μM/mm emerges within the swarm which is comparable to the gradients which bacterial species such as *E. coli* (Jeon et al., 2009)(Diao et al., 2006) have been reported to sense. We found that this gradient of the signalling molecule concentration accurately predicts the future direction of the swarm. Specifically, the swarm will move in the direction of the steepest negative gradient although inertia and stochastic effects would induce some deviations around this expected direction. These effects are not included in this model currently. However, this model serves to demonstrate that for reasonably realistic parameter values, measurable concentration differences of the signalling molecule emerge within the swarm which regulate the direction of the swarming tendril. In the case of sensing of sister tendrils, we see that μM/mm gradients appear typically only in the smaller swarm tip which then changes its direction of motion [Figure 3(a)(f)]. In the case of inert obstacles, we again see that the presence of a reflecting boundary results in measurable gradients which predict the future direction of the swarm [Figure 4]. For this case, we find the strongest argument for a bacterium's requirement of membership in a swarm because measurable gradients from inert reflecting boundaries will only form if the source strength is large enough. The large source strength required is provided by the swarm whereas individual bacterium would never be able to produce it on its own.

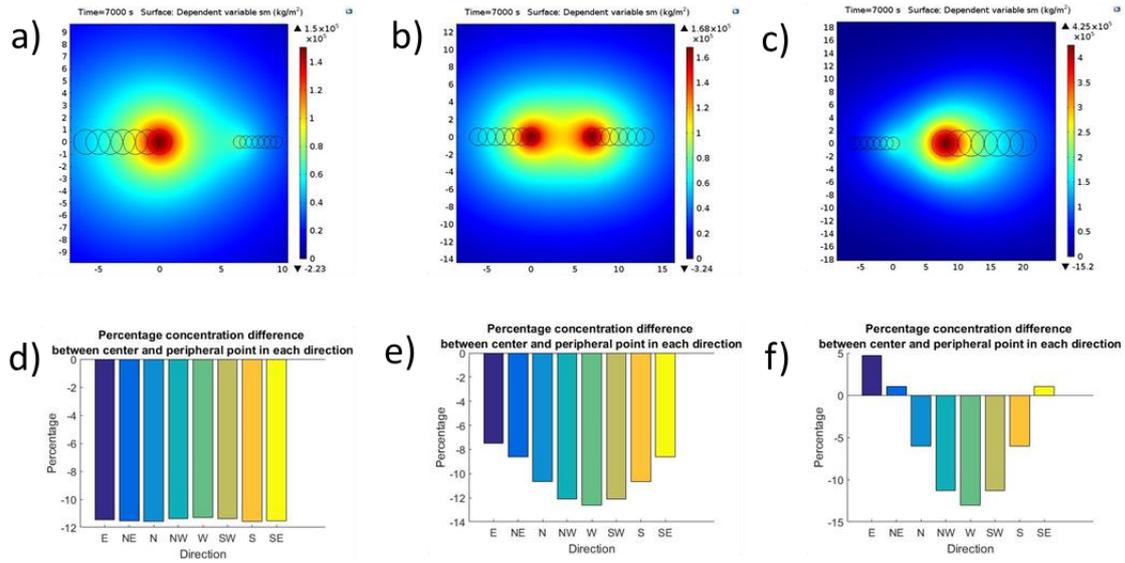

*Figure. 3, Signalling molecule gradient which emerges within the swarm tip when encountered by other approaching swarm tips of different sizes at a separation of 5mm.*

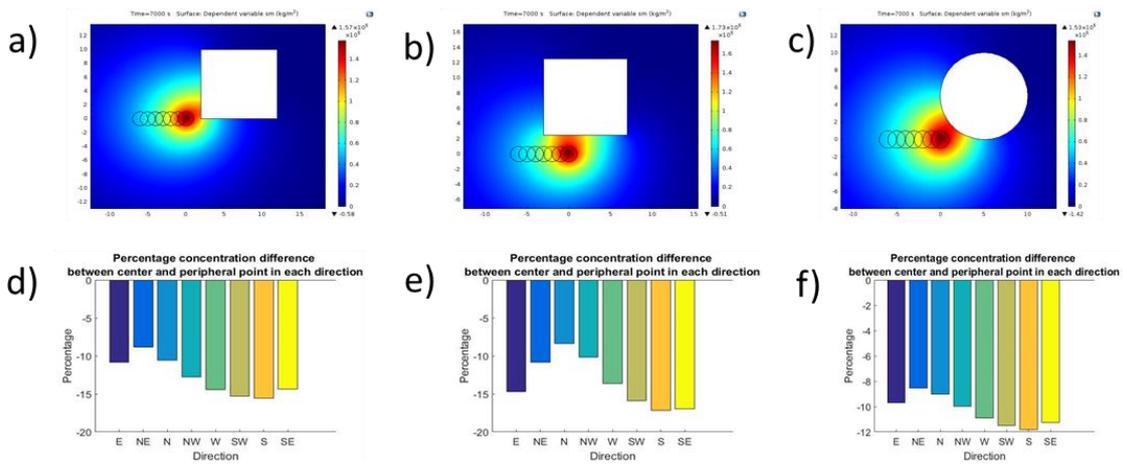

*Figure 4, Signalling molecule gradient which emerges within the swarm tip due to an inert obstacle at a separation of around 1 mm.*

The model discussed above is a coarse-grained model which assures us of the strong likelihood of measurable gradients of a signalling molecule forming within the swarm. Ultimately, we would like to understand collective spatial awareness of the swarm emerging from individual behaviour and cell-cell interactions including its inherent stochasticity. Insights derived from such studies can guide robust distributed control strategies for future robotic swarms (Rubenstein et al., 2012)] and other collective sensing phenomena. We constructed a multi-agent model [Wilensky, U. (1999). NetLogo] based on two-component signalling system. The two signalling molecules produced by the agents have different diffusion constants with each of them either invoking an attracting or repelling response among the agents. The spatial distribution of the signalling molecules govern the behaviour of the agent [See SI Section 8 for a detailed description of the multi-agent model], which spontaneously shows the emergence of branching as observed in experiments [Figure 5]. The agent based system also exhibited spontaneous emergence of spatial awareness similar to the biological system, most notably, the detection of boundaries (analogous to detection of inert obstacles presenting non-diffusive reflective boundaries) and consequent retraction of the advancing swarm tendril as observed in experiments. The spontaneous emergence of these features in the simulations suggest that the remarkable spatial awareness seen in the bacterial swarm may arise from simple attractive-repulsive interactions between bacteria which indirectly leads to the effective computation of the global gradient predicted by the fluid dynamic model.

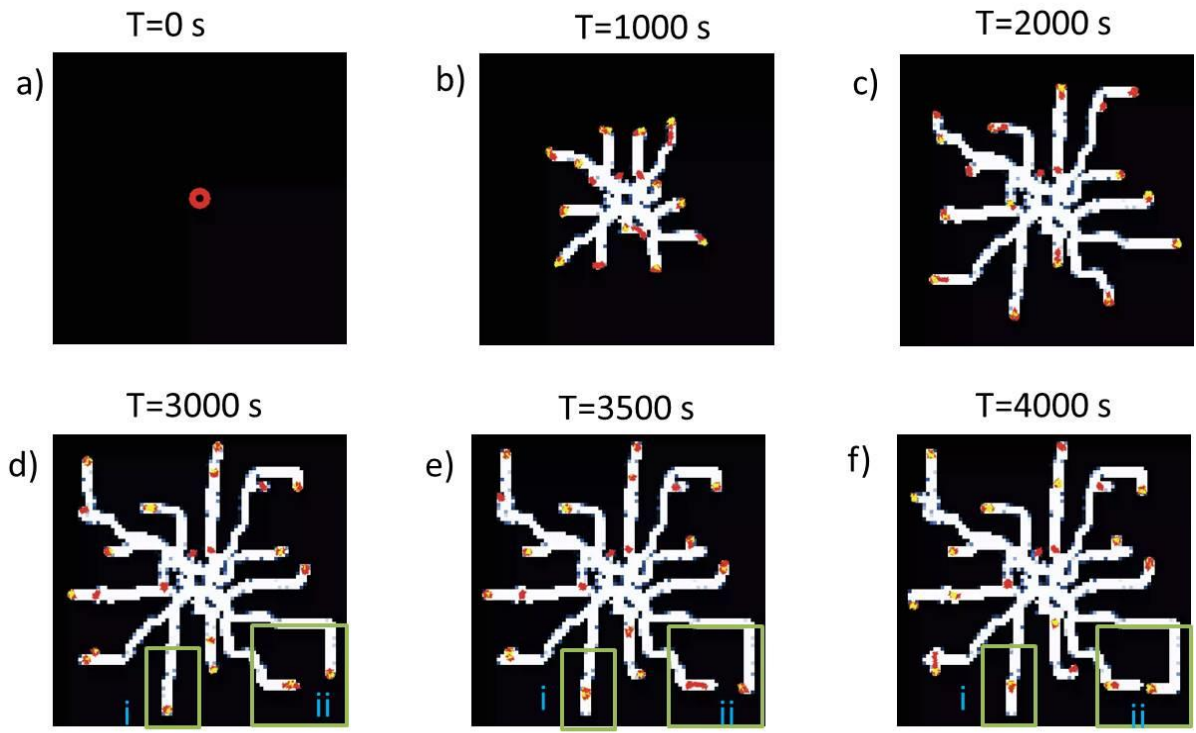

*Figure. 5, Illustration of agent based simulation showing branched pattern and retraction from boundary [ Red pixels = Active swarmers in state 0, Yellow pixels = Active swarmers in state 1, White pixels = water film formed or colony spread (See SI section 7 for more details)]. Regions i and ii highlights the retraction phenomena from the boundary and other tendril respectively.*

**Discussion and open questions**

Detailed spatial awareness expectedly increases the fitness of an organism to survive a complex environment. In this article, we showed examples of a primitive organism displaying sophisticated spatial awareness such as long-range sensing of an inert obstacle. The response of the swarm to these perturbations is coherent and highly coordinated unlike previous descriptions of sensing neighbouring swarming colonies by lethal secretions. The role of a relatively fast diffusing signalling molecule is expected and strongly suggested by the mathematical models. In particular, the coarse-grain model provides strong evidence to the fact that measurable differences in the concentration of signalling molecules can arise from our expectations on the model parameters which will induce difference in behaviour in individual bacteria at the swarm boundaries. Indeed, a single cell resolved, multi-agent model based on signalling molecule concentration dependent attractive-repulsive potential exhibits spontaneous emergence of branching and some aspects of spatial awareness observed in the actual biological system. Presently the agreement between experiments and models is largely qualitative. The patterns observed in the multi-agent simulations do not resemble the patterns observed in the experiments indicating that the present model is not complete. Another open question is how the swarming tendril avoids self-inhibition from its own secretions while being able to inhibit the advancement of a neighbouring tendril at a much longer distance away. Although these questions are still open, this work presents some essential ingredients likely to lead to a quantitatively rigorous agent based model for PA swarming which can reproduce not only the pattern formation aspects but also its collective sensing abilities of the swarm. Such a model would also enable one to study robustness of the collective behaviour in the presence of defective individuals and other perturbations. Insights related to robust behaviour would be of exceptional value to the emerging field of swarm robotics from the perspective of robust decentralized control.

## Methods

### Swarming Motility Assay

For swarming assay, we used Peptone growth medium (PGM). Composition of PGM 0.6% agar plates are 6 grams of bacteriological agar (Bacto agar), 3.2 grams of peptone, and 3 grams of sodium chloride (NaCl) added in 1 litre of distilled water. The medium was autoclaved at 121°C for 30 minutes. After autoclaving the media, 1 mL of 1M $CaCl_2$ (Calcium chloride), 1 mL of 1 M $MgSO_4$ (Magnesium sulphate), 25 ml of 1M $KPO_4$ and 1 mL of 5 mg/mL cholesterol were added into the medium and mixed properly. 25 mL PGM were poured in each 90mm Petri-plates and allowed them to solidify at room temperature (RT) for a half an hour under the laminar hood flow with the lid opened. And all the plates were kept at room temperature for 16-18 hours for further drying.

### Swarming

2ul of a planktonic culture of Wild type PA or flgM mutant with OD >2.8 is inoculated at the centre of 90 mm petri-plate containing PGM-0.6% agar. The wild type PA14 forms the pattern as shown in Figure 1 a) over a period of 24 hours in a 90 mm petri-plate. FlgM is a transposon insertion mutant in the flgM gene of PA14 and is part of *P. aeruginosa* transposon insertion library (Liberati et al., 2006).

### Preparation of PDMS (Poly DiMethyl Siloxane) obstacle

We have used Sylgard 184 from Dow Corning. It has two parts: an elastomer part and the curing agent. The two parts i.e. elastomer: curing agent is mixed in the ratio of 10:1. This mixture is stirred well. The air bubble cause due to stirring is removed by degassing the PDMS mixture in a desiccator connected to vacuum pump. An acrylic template is made to obtain different shapes of the obstacle. The air bubble free PDMS mixture is then poured into the template and cured for 12 hours. The cured PDMS solidifies and is removed from the acrylic template. These PDMS obstacles are then sterilised in autoclave.

The sterilised obstacle blocks are placed in an appropriate position in the petri-plate. The nutrient agar is then poured around the obstacle such that the obstacle is half immersed in the nutrient agar while held intact in its original position. The nutrient agar with the obstacle is allowed to dry under the laminar hood.

## Acknowledgements

We gratefully acknowledge Robert Bosch Centre for Cyber Physical Systems at Indian institute of Science, Bangalore, India for funding this research. We also acknowledge the use of facilities at Centre for Nano Science and Engineering, Indian Institute of Science, Bangalore, India.

# Supplementary information

## Section 1

**Different swarming patterns of different bacteria- sparsity compactness analysis**

Each bacterial species produces a unique swarming pattern. To characterize the uniqueness of each pattern we define two factors: Sparsity and Compactness.

Compactness helps in differentiating a branched pattern from a circularly expanding one. Compactness is defined as the ratio of the area to the square of the perimeter. Circle has compactness equal to 1. A highly branched pattern has higher perimeter for a given area hence compactness is very low.

Compactness

$$C = \frac{4*pi*Area}{Perimeter^2}$$

Sparsity factor shows how well separated the branches are. The closely packed branches are less sparse.

Sparsity

$$S = 1 - \frac{Area\ of\ pattern}{Area\ of\ minimum\ bounding\ circle}$$

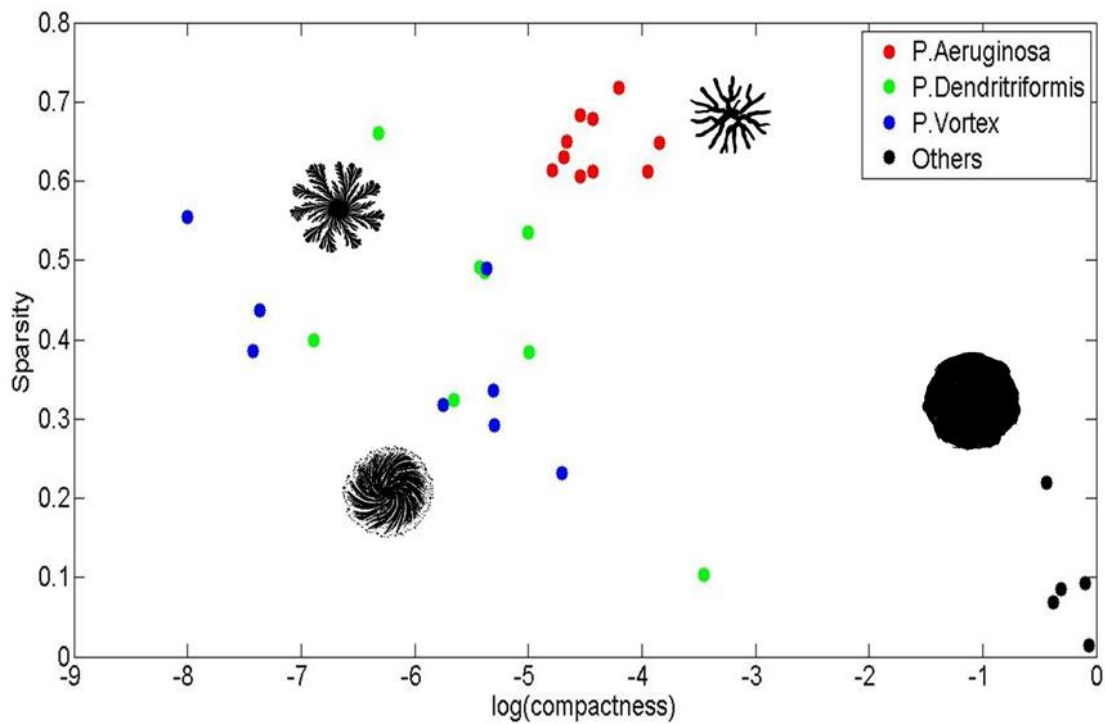

*Figure S1. Sparsity vs. log(compactness) for swarming patterns of different bacteria*

The graph shows *pseudomonas aeruginosa* clusters are well separated from the patterns of the rest of swarming bacteria. This indicates the uniqueness of the pattern and the need for a unique model to describe the pattern as the existing models that explain the other patterns may not hold good.

To characterise the swarming pattern of the *Pseudomonas aeruginosa*, we calculate the branch length, branch width and branch angles. The statistics of the pattern parameters have been summarised in the Figure S2.

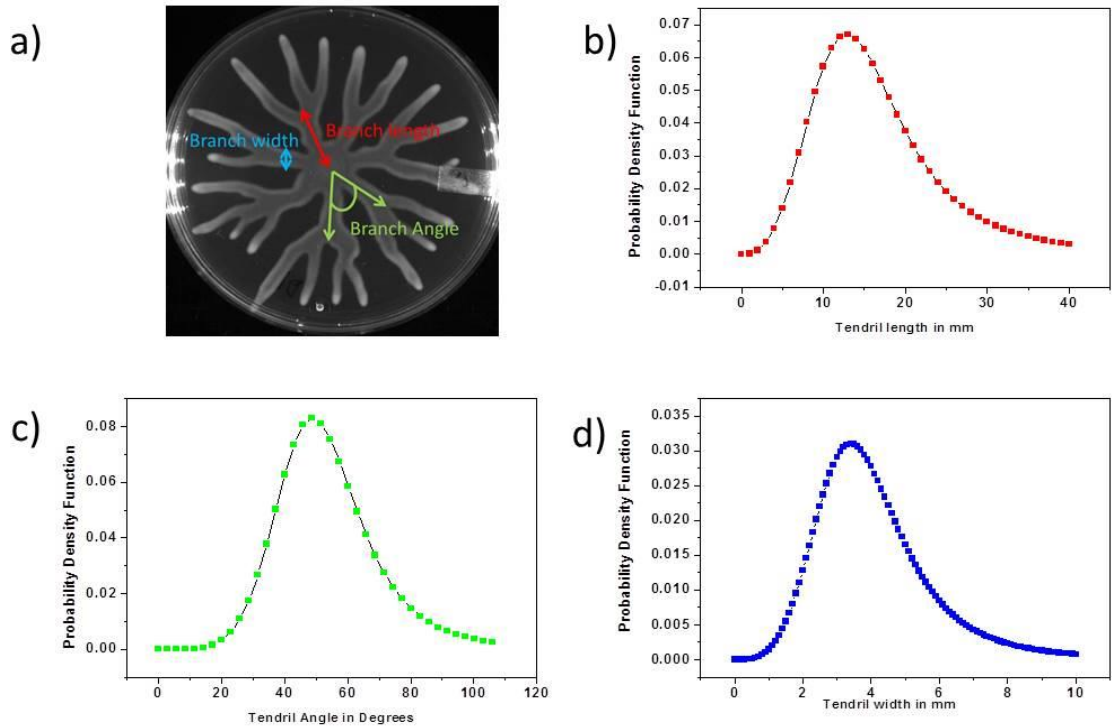

*Figure S2. a) Swarming pattern of PA on a PGM-0.6% agar with labelled branch parameters. Statistics Probability density function of different branch parameters have been obtained with b) mean tendril length is 16.85 mm c) mean tendril angle is 54.04 degrees and d) mean tendril width is 4.4 mm.*

The tip of the branch has been tracked in the time lapse video recorded over 24 hours. The velocity of the branch TIP has been calculated for 10 such branches and the average was found to be about 55.48 µm/min

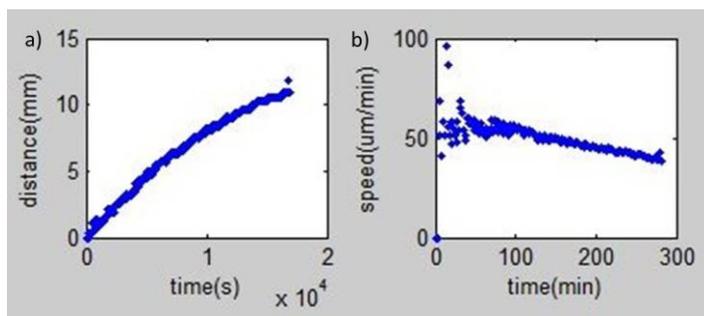

*Figure S3., a) distance covered by the tip of the tendril at different time points from the starting point, b) speed of the tip of the tendril is almost constant at 50 µm/min*

# Section 2

## Awareness of the Presence of Sister Tendrils

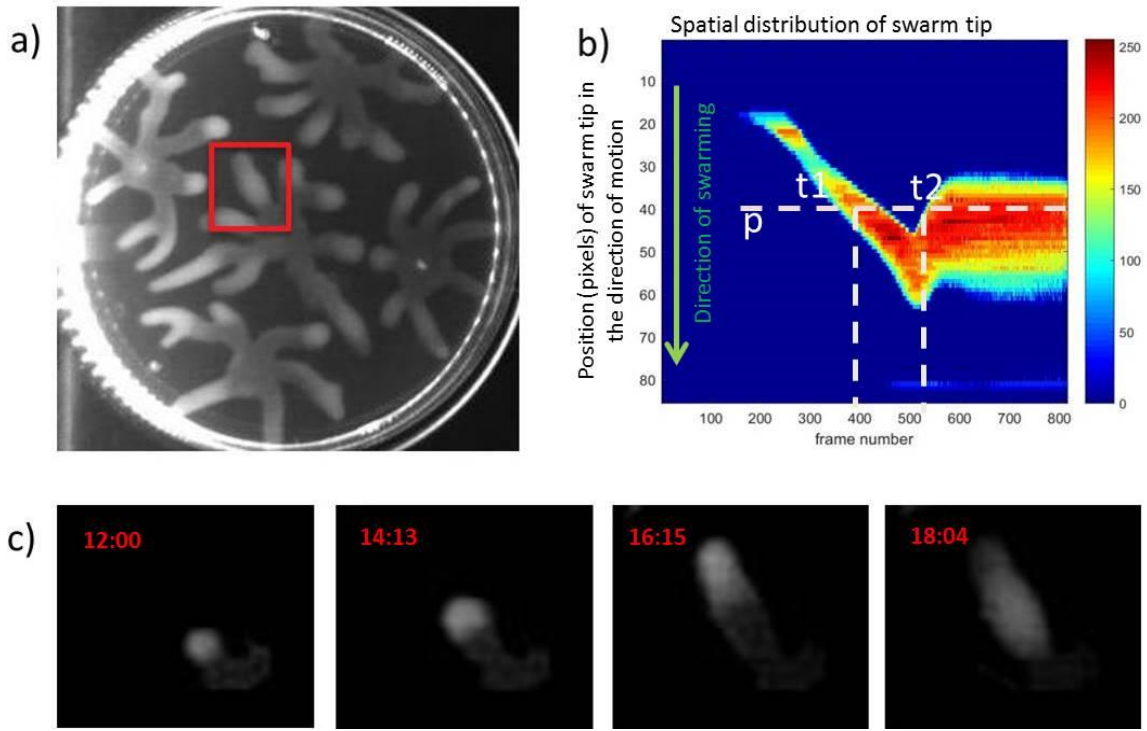

*Figure S4., a) Swarming pattern respond to a sister colony on the same petri-plate, b) swarm tip position (3.35 pixels/mm) and density with increasing time(0.67 frames/min)(t1 and t2 are time points at which the swarm tip passes a point ' p' along the tendril, one during forward swarming and the other during retraction ), c) example of swarm tip progression and retraction with time*

Along with awareness of the edge, the tendrils can sense the sister colonies and the branches of their own colony. The Figure S4 a) shows five colonies on the same petri-plate. They show two types of behaviours, either change in direction to avoid colliding with the other branches or active cells stop and retract when they find no other suitable direction to progress.

When the wild type is allowed to swarm on the same plate as a mutant like flgM which do not swarm, we again see interesting behaviour of avoidance as illustrated in the Figure S5 .[ See video PA_mutant_interaction.avi]

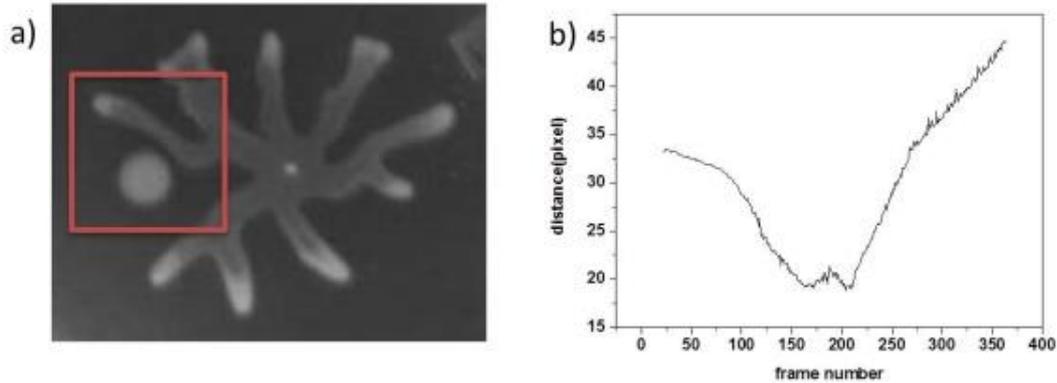

*Figure S5. a) Swarming pattern respond to a non swarming mutant on the same petri-plate b) variation of distance between swam tip and the center of the non swarming mutant with time (0.56 frames/min).*

**Section 3**

The video shows wild type PA swarming on a 90mm petri-plate and the swarm tip retracts from the edge of the petri-plate. [See video Edge_retraction.avi]

**Section 4**

**Obstacle made of different materials**

To test the independence of the avoidance behaviour to the material used as the obstacle, we have used obstacles made of glass and steel. We see the similar phenomena for the obstacles made of glass and steel, indicating the material independence of the avoidance behaviour. This strengthens our hypothesis of the obstacle being just a reflecting boundary and not a source of any signalling molecule that could be sensed by the bacteria.

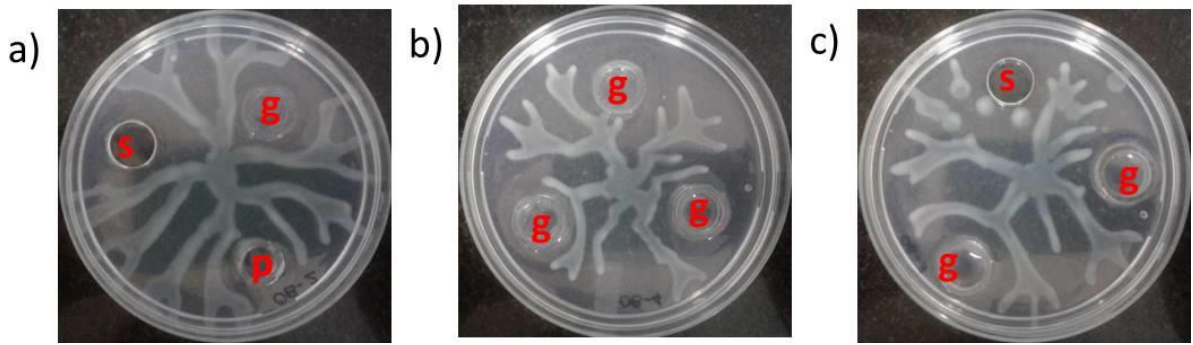

*Figure S6. Swarming patterns in the presence of obstacle made of a) steel(s), glass(g), PDMS(p), b) all of glass(g), c) two of them made of glass(g) and one of steel(s)*

**Obstacle videos**

The time lapse videos of the wild type PA in PGM media and M9 media in the presence of obstacles made of PDMS.

[See videos Obstacle_PGMmedia.avi , Obstacle_M9media.avi]

**Section 5**

**Asymmetry Analysis to quantify perturbation to swarm pattern**

To test the effect of the obstacle on the swarm pattern of the wild type, we take wildtype-flgM plate as a positive control to get a baseline for a definitely biased swarm pattern and a normal swarming plate as negative control to get a baseline for unbiased swarm pattern. We expect the system with an inert obstacle to lie in between these extremes.

The average size of the obstacles is comparable to the mutant spread on the plate. This size is small compared to the area available in-between branches. Since the pattern is quite sparse, we need to rule out the possibility that the branch might have avoided the obstacle by chance. So in our analysis, we translate the obstacle along the pattern and check for regions where the obstacle might occur. Each pixel position represents the centroid position of the obstacle and the intensity of the pixel represent the degree of overlap of the obstacle with the pattern. Yellow pixel represents the position of maximum overlap with the pattern and the dark blue would indicate that obstacle would fit perfectly in the position and would not intersect the swarm pattern. We analysed about 120 pattern with different shapes of obstacles like triangle, rectangle and a circle .

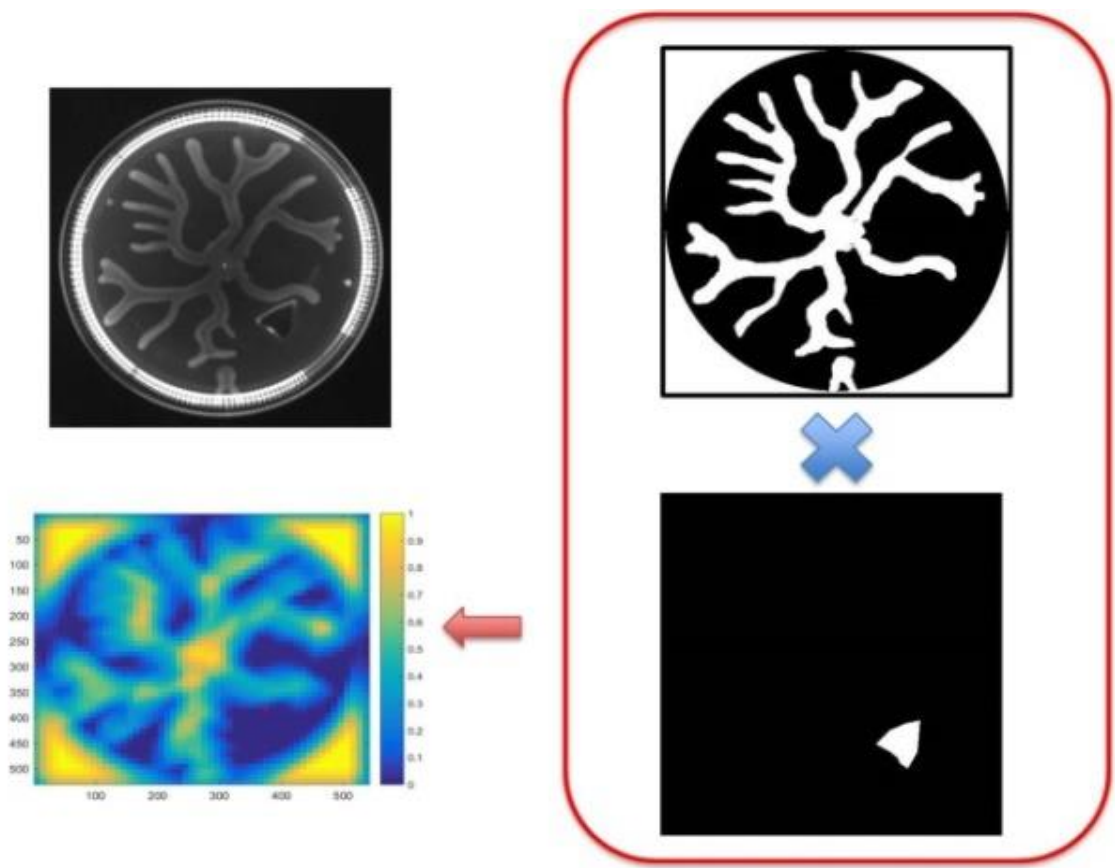

*Figure S7.  Obstacle translation method to generate a heat map representing the suitable positions the obstacle can occupy without touching the swarm pattern*

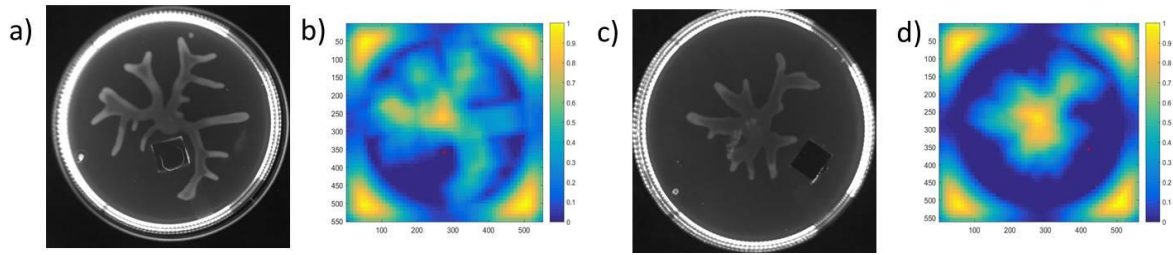

*Figure S8. (a),(b) Example of case of most probable detection and its corresponding heat map (c),(d) Example of Undecidable cases and its corresponding heat map*

The heat maps of 120 swarms with obstacles show open positions on each petri-plate that a swarm could occupy without touching the tendril(s). This allowed us to categorise the obstacle sensing into 2 distinct patterns- 1) Most probable detection- The obstacle is present in the largest open position in the swarm pattern. This indicates that the presence of the obstacle somehow perturbed the branches to move away from the obstacle leading to shift in patterns, 2) Undecidable case - the swarm does not progress till the end of the petri-plate, hence it is difficult to say if the swarm detected the obstacle or not.

In the presence of the obstacle, we see that about 80 % of the plates showed most probable detection. This indicates the presence of obstacle indeed perturbs the PA swarms and their behaviour. This can be a new example of the social communication and decision making that PA is known to exhibit.

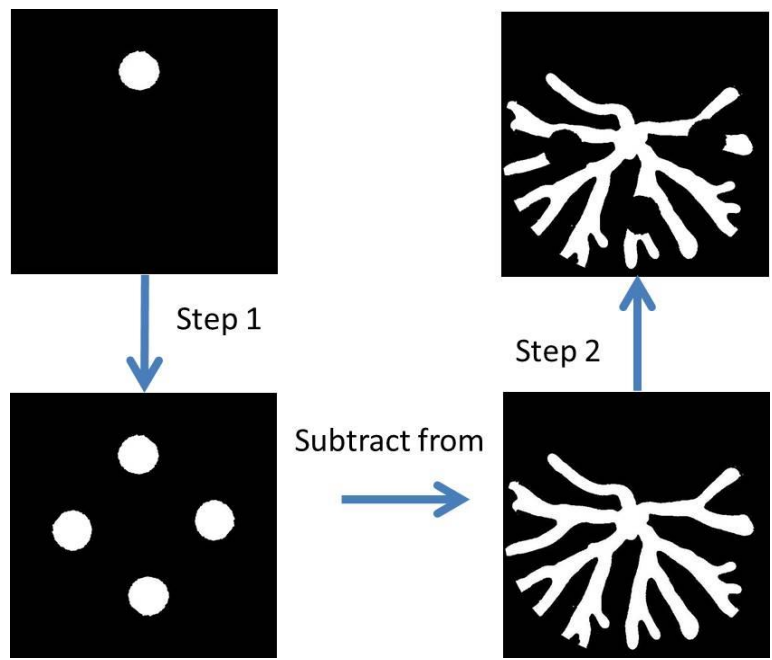

*Figure S9: step1: Replicating the obstacle in all four quadrants to obtain a symmetric distribution step2: Subtracting symmetric obstacle distribution from the perturbed swarm pattern induced by the obstacle*

To quantify the obstacle induced perturbation of the swarm, the symmetry breaking of the pattern needs to be quantified. The branch distribution in the case of a negative control (no obstacle) can be assumed to be unbiased hence symmetric. But the symmetry was statistically (80 out of 120 plates) affected in the tendrils around the obstacle. The presence of obstacle may create asymmetry in the nutrient distribution thus inducing change/asymmetry in the swarm behaviour and pattern. To negate this asymmetry, we digitally created images of obstacles in four quadrants(Figure S9 step1) and subtract the obstacle region from the pattern region(Figure S9 step2). This will remove the asymmetry in the pattern due to asymmetry in the nutrient distribution itself. Now the pattern obtained after subtracting the digitally created obstacle distribution is subjected to symmetry analysis (Figure S9).

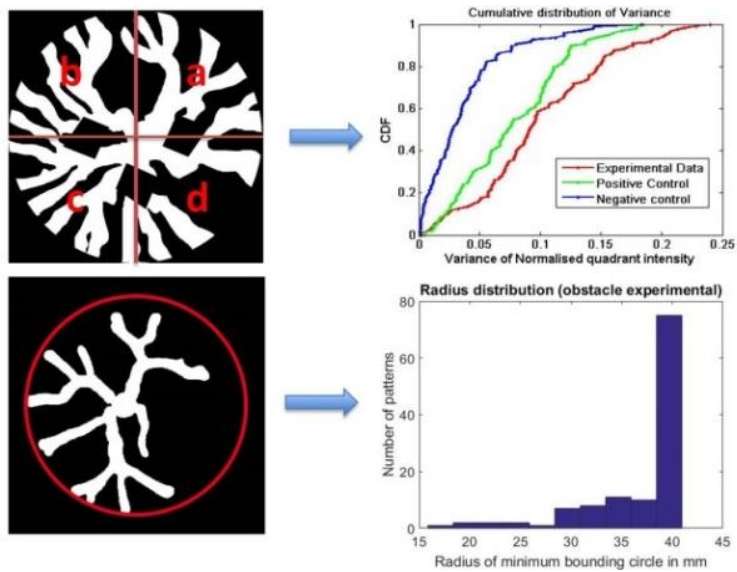

*Figure S10: a) Inter-quadrant asymmetry analysis of swarms with obstacles. b) Minimum bounding circle analysis of swarms with obstacles.*

The pattern is divided into four quadrants. The white pixels represent the bacterial swarm and black pixels represent the part of the image excluding the swarm. The number of the white pixels in each quadrant indicates the extent of the colony spread in that quadrant. By calculating the variance of the number of white pixel distribution over the four quadrants, extent of symmetry in the pattern can be quantified. Higher the inter-quadrant variance, higher is the difference in swarm area distribution among quadrants and hence higher the asymmetry. Figure S10 shows the mean variance of the pattern in different quadrants. The mean variance is clearly higher for obstacle experiments and is comparable with the experiments with flgM mutants on the same plate. This indicates that there is asymmetry in the pattern when physically or chemically perturbed.

A control plate of swarming PA has tendrils reaching the edge of the petri-plate in 24 hours. For the asymmetry argument to hold, the swarms should have reached the periphery of the petri-plate in the direction of no perturbation. To verify this, a minimum bounding circle for each plate inoculated with the bacteria at the centre is considered. This can be quantified by measuring the minimum radius of the circle with its centre fixed at the point of inoculation (which is usually the center of the plate) and can completely bound the pattern. The radius of the minimum bounding circle is equal to the radius of the petri-plate in control plates. The petri-plate considered for the experiments are of 40 mm radius and from the histogram more than 75% of the plates have a minimum bounding radius at 40 mm thus confirming the pattern swarm till the end of the petri-plate.

## Section 6

**Experimental parameters**

In a swarm on PGM-0.6% agar, tendrils start at the swarm centre and move away with an average speed of 1 µm/s after some lag period. The mass of a bacteria is approx. 1pg. Assuming, it produces signalling molecules equal to its body weight in its complete life time. The cell division/binary fission time of the bacteria is about 60- 90 minutes in minimal media (Badalà et al.,2008).we measure the cell density at the tendril tip to be 1.5e6 – 2e6 cell/mm^2. The rate of production of signalling molecule is thus

$$rate\ of\ signalling\ molecule\ production \sim \frac{mass\ produced * cell\ density}{generation\ time}$$

$$= \frac{1 * 10^{-12} * 2 * 10^6}{60 * 60}$$

$$= 0.555\ ng/mm^2.s$$

For simulation, secretion rate is taken to be 1ng/mm^2.s

The active swarm cells have been observed to change direction by sensing the presence of the other branch from distances as far as 5mm. Assuming the communication is happening only based on the gradients of the self-produced signalling molecules, the long range sensing observed in the experiments is possible only if the signalling molecules diffused faster than the distance covered by the moving sources. This constraint can be represented mathematically as

$$\frac{L^2}{D} \ll \frac{L}{v}$$

L= length at which sensing happens

D=diffusion constant of signalling molecule

V= velocity of swarm tip

Taking L=5mm and velocity of about 1 µm/s (experimentally determined)

$$D \gg 5 * 10^{-9}\ m^2/s$$

Diffusion constant of D=1e-8 m^2/s is considered for simulations. This diffusion constant implies a small signalling molecule with the Stokes radius of 25.5 pm (This corresponds to the radius of a hydrogen atom assuming diffusion is happening in water).

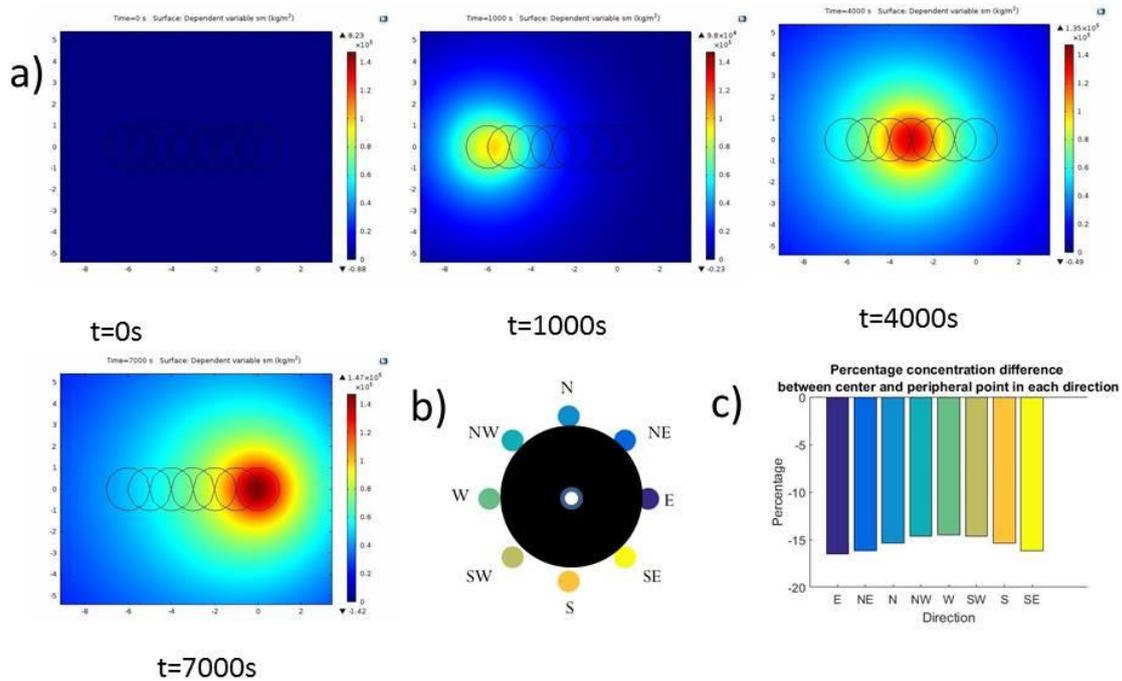

*Figure S11: a) distribution of signalling molecules at different time stamps b) probe points of the swarm tip c) Percentage concentration difference between the center and peripheral point of the swarmtip in each direction*

The simulation involves the swarm tip of about 1mm in radius. It moves about 1mm in 1000s which is equivalent to 1μm/s (experimentally, each position within the swarm tip produces 1ng/mm^2.s of the signalling molecules).

The simulation results indicate that the concentration difference between pairs of orthogonal points shown in Figure S11, assuming an agar height of about 5mm, ranges from 0.2 mg/L – 2mg/L. If the signalling molecule was to be Rhamnolipid (molecular weight of about 650 g/mol) , we get the concentration gradient in the range of 0.3 μM/mm – 3 μM/mm.

**Section 7**

**Multi-agent Modelling**

**Swarm states**

The model considers 500 active swarmers initially at the center of the 81x81 patch as shown in the Figure S12 a). All the patches initially contain no chemical or water in them. Quorum sensing is necessary for the active swarmers to know if the local density of the swarmers has reached the critical density enough to produce sufficient surfactant to ease their movement across the surface. The active swarmers in the model are hence in 2 states.

State 0: Swarmers know that the quorum has not been reached

State 1: Swarmers know that the quorum has been reached and change their behaviour

**Quorum signal(attractant) production**

Initially all the active swarmers are in state 0 and each headed in random directions. Each swarmer in state 0 produces quorum signalling molecule at rate 0.2 mass units/second (1 second is equivalent to a tick in simulations) and looks for the direction in which the quorum signal molecule is highest. The search is restricted to the patches defined by check length l (set to 1 in our simulation) and check angle theta (set to 45 degree). The check length l is measured from the current position of the swarmer and check angle theta is measured from the current direction of heading as shown in Figure S12 b). The direction of swarmer is updated to the best among the patches after the check process. If none of the left or right patches are better than the patch ahead, the swarmer continues to move in the original headed direction. The swarmer checks the current patch value of quorum molecule at each step and changes its state accordingly. Let Cqi be the value of quorum signalling molecule in patch in which the i$^{th}$ swarmer is present and Si be the state variable of the i$^{th}$ swarmer

If (Cqi < Tq)  Si =0 else Si =1

Where Tq = quorum threshold

**Repellent production**

If the swarmers are in quorum (state 1), they produce surfactant. As a result a thin water film is formed in the current patch. Each swarmer in state 1 increases the water content of its current patch position by 0.1 mass-units/second.   Along with the ability to create water film, they also produce repellent. Swarmers produce repellent at a rate of 0.2 mass-units/second. Along with producing repellent, they get away from regions with high concentration of the repellent. They use the same checking mechanism discussed above and choses the minimum repellent direction for the next time step.

**Water edge detection**

The swarmers move to the edge of the water film created by them to expand their colony into niche nutrient rich zones. They move with different speeds depending on the presence of water in the patch where they are located. If the patch has water, it moves 0.05 step length per time and if the patch has no water, it moves 0.01 step length per time.

The direction of the motion is decided by the current headed direction which gets updated depending on the swarmer's state and the values of quorum signal, repellents and water around the swarmer.

**Diffusion**

The quorum signal and repellent molecule diffuse to the neighbouring patch positions using rules described by the following equations

$$c(x,y,t) = c(x,y,t-1) - \frac{n * D * c(x,y,t-1)}{8} + \sum_{i=1}^{n} D * c_i(x,y,t-1)/8$$

Where c(x,y,t) is the concentration of a chemical in patch (x,y) at time t

n is the number of neighbours to a patch at position (x,y)

ci(x,y,t-1) is the concentration of the chemical in the ith neighbour of the patch (x,y)

D is the diffusion constant of the chemical

The value of D of quorum signal is 0.2 and D of repellent is 0.8. We observed that the diffusion constant of the quorum signal had to be less than that of repellent to observe branching patterns in the simulations, qualitatively similar to those observed in experiments.

**Evaporation**

The quorum signal molecule evaporates and reduces its concentration at each time step by 10%.

**Noise**

Noise is added to the final direction chosen by the swarmer. It is in the range -45 to +45 degrees.

The video of the simulation of the swarm based on the above rules is available at the following link. [See video swarm_simulation.avi]

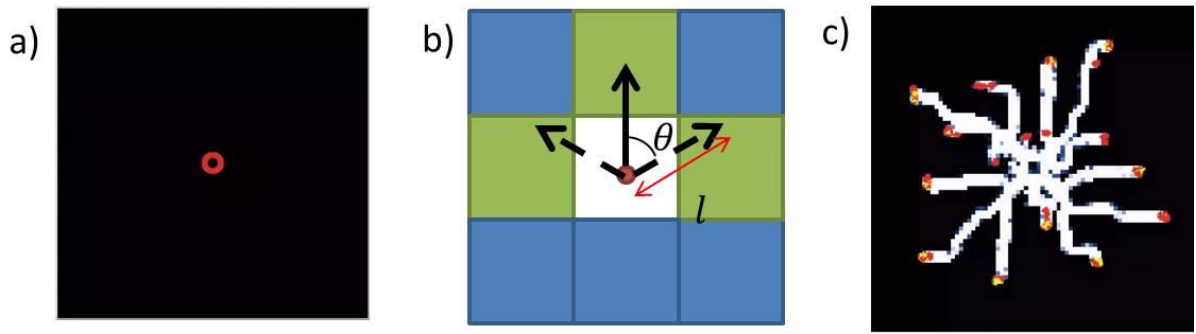

*Figure S12., a) Initial state of the active swarmers, b) Search strategy (solid black line – current headed direction of the swarmer, green patch – check patches for next direction of movement), c) Simulated Branch pattern with agent represented in red or yellow depending on the state being 0 or 1 and water in white.*